\documentclass[conference]{IEEEtran}
\IEEEoverridecommandlockouts

\usepackage{graphicx}
\usepackage{comment}
\usepackage{amsmath}
\usepackage{amssymb}
\usepackage{array}
\usepackage{enumitem}
\usepackage[colorlinks,urlcolor=blue,linkcolor=blue,citecolor=blue]{hyperref}

\usepackage{hyperref}
\hypersetup{ 
 colorlinks=true, 
 linkcolor=blue, 
 filecolor=blue, 
 citecolor = blue, 
 urlcolor=blue, 
} 

\begin{document}

\title{Network-wide Quantum Key Distribution with Onion Routing Relay}

\author{\IEEEauthorblockN{Pedro Otero-García, David Pérez-Castro, Manuel Fernández-Veiga, Ana Fernández-Vilas 
}
\thanks{This work was supported under the grant TED–2021–130369B–C31 funded by MICIU/AEI/
10.13039/501100011033 and by the “European Union NextGenerationEU/PRTR” and the grant  PID2020–113795RB–C32 funded by MICIU/AEI/10.13039/501100011033.}

\IEEEauthorblockA{atlanTTic, Universidade de Vigo \\
\{pedro.otero,dperezcastro,mveiga,avilas\}@det.uvigo.es}}

\maketitle

\begin{abstract}
The advancement of quantum computing threatens classical cryptographic methods, necessitating the development of secure quantum key distribution (QKD) solutions for QKD Networks (QKDN). In this paper, a novel key distribution protocol, Onion Routing Relay (ORR), that integrates onion routing (OR) with post-quantum cryptography (PQC) in a key-relay (KR) model is evaluated for QKDNs. This approach increases the security by enhancing confidentiality, integrity, authenticity, and anonymity in quantum-secure communications. By employing PQC-based encapsulation, ORR pretends to avoid the security risks posed by intermediate malicious nodes and ensures end-to-end security. Results show that the performance of the ORR model, against current key-relay (KR) and trusted-node (TN) approaches, demonstrating its feasibility and applicability in high-security environments maintaining a consistent Quality of Service (QoS). The results show that while ORR incurs higher encryption overhead, it provides substantial security improvements without significantly impacting the overall key distribution time.
\end{abstract}

\section{Introduction \label{sec:intro}}

The development of quantum computing is rapidly evolving~\cite{aasen2025}, and with recent advancements in quantum computation~\cite{acharya2024},  algorithms such as Shor's~\cite{shor1999} and Grover's~\cite{grover1996} will be imminently implemented. As the majority of current cryptographic systems are vulnerable to quantum attacks, an attacker with quantum computing capabilities would jeopardize all classical computer security.

In particular, public-key algorithms, such as RSA~\cite{rsa1978}, Diffie-Hellman~\cite{diffie1976} and elliptic curve-based algorithms (ECC)~\cite{koblitz2000}, will lose effectiveness because they are based on limited computational assumptions about the potential attacker, which are broken by quantum computers. For instance, Shor's algorithm can factor large numbers in polynomial time, completely breaking the security of RSA, for a sufficiently powerful Quantum Processing Unit (QPU). Similarly, the discrete logarithm problem, the basis of ECC and Diffie-Hellman, also becomes ineffective in front of quantum attacks.

In the case of symmetric encryption algorithms such as AES~\cite{aes2001} and cryptographic hash functions such as SHA-2 and SHA-3~\cite{nist-hash}, the quantum threat is also relevant, although less severe. Grover's algorithm can reduce the time required of a brute force attack against a symmetric encryption algorithm from \(2^n\) to \(2^{n/2}\), which implies that, to maintain the same level of current security, the size of the keys would have to be doubled. In other words, to achieve the current level of security provided by AES-128, AES-256 would need to be employed, meaning that depending on the specific implementation process, encryption times could double. 

Given the forthcoming scenario, the scientific community has put intensive efforts to develop cryptographic solutions resistant to quantum attacks. There have been two main approaches: quantum cryptography (QC); which relies on the physical properties of quantum systems to guarantee communication security, and post-quantum cryptography (PQC); which comprises the search of algorithms capable of resisting the attacks of QPUs. While both approaches have its merits, a complete and reliable solution is still lacking. 

As stated before, QC focuses on designing quantum resistant processes and systems by taking advantage of the nature of quantum systems to guarantee security. Notably, Quantum Key Distribution (QKD), allows secure key exchange between end users by using quantum signals. There are two distinct protocol groups: Prepare and measure protocols; which base its security in quantum superposition, with BB84~\cite{bennett1984} being the most prominent and widely used, and entanglement-based protocols; where E91~\cite{ekert1991} and BBM92~\cite{bennett1992} are highlighted. 

The main advantage over other key agreement protocols is that the shared key is Information Theoretically Secure (ITS), while the main drawback of this collection of QKD solutions is the actual implementation of the technologies. Currently, no commercial solution for quantum repeaters is available, heavily limiting the communications as a consequence of the sensibility of quantum hardware and the fragility of quantum signal transmission, limited to approximately 100 km point to point quantum transmission in fiber~\cite{stanley2022}. Although critical improvements~\cite{main2025} are being made and milestones are constantly reached, complex QKDNs are still not currently feasible in conjunction with classical infrastructure and require specific hardware. The high cost associated with implementing this new dedicated structure presents a significant barrier, making large-scale networks unfeasible during the Noisy Intermediate-Scale Quantum (NISQ) era.

On the other hand, PQC allows for a quantum resistant solution which is easier to adopt, as no specialized hardware is needed to be implemented and offers, in principle, as much security in communications. Therefore, while the security of QC resides in the very nature of quantum mechanics and no assumptions on the computational power of the attacker need to be made, PQC algorithms are based in mathematically complex problems which could be broken by a powerful enough attacker. For this reason, significant effort is directed by institutions to standardize PQC algorithms. The National Institute of Standards and Technology (NIST) is leading this initiative, and algorithms have already been announced as new standards:  CRYSTALS-Kyber~\cite{nist-fips203} (Kyber), a public-key encryption scheme; CRYSTALS-Dilithium~\cite{nist-fips204} (Dilithium), a digital signature algorithm; and SLH-DSA~\cite{nist-fips205}, a hash-based signature protocol. 

Using cost-effectiveness as the key metric, PQC has major advantages over QC, since its implementation is more immediate, versatile and economical. However, despite the hardware limitations of QC, QKD remains the only option that has been proven to be completely secure against quantum attacks, and it will be the only solution that will guarantee ITS for critical infrastructures in short and long term. The two current approaches that try to provide a solution to the scalability problem of QKDNs are key-relay (KR)~\cite{elliott2002,collins2005, james2023} and trusted-node (TN)~\cite{itu-y3803} (or central Key Management System (KMS)). Nevertheless, both models do not respect the end-to-end confidentiality of the key by the intermediate nodes. Additionally, as current QC technology cannot handle the scalability issues, a hybrid solution (QC and PQC) is mandatory to guarantee unconditional point-to-point security in a QKDN.

In this context, this article evaluates a secure key distribution protocol between any two nodes inside a network that do not have a quantum channel to connect them, using the KR based key distribution model, complemented with the encapsulation philosophy of Onion Routing~\cite{goldschlag1999} (OR) and post-quantum techniques. This approach maintains the principles of Confidentiality, Integrity and Authenticity (CIA) in a secure network while incorporating the communication anonymity assured by OR, to maintain the highest possible degree of security and quality of service (QoS).

The rest of the document is organized as follows: In the following section~\ref{sec:related}, the article reviews existing QKDN security models, discussing KR and TN approaches, compares them with the ORR model. Section~\ref{sec: tech-bg} provides essential details on KR and TN in QKDNs, Onion Routing principles, and the role of Kyber and Dilithium PQC algorithms. Meanwhile Section~\ref{sec:model} introduces the ORR model, explaining its security enhancements and comparing it to traditional KR and TN methods. Section~\ref{sec:tests} details the experimental setup, cryptographic tools, and performance metrics used to evaluate ORR’s feasibility. The results obtained are presented in Section~\ref{sec:results} where a comparative performance analysis is shown, focusing on encryption time, key distribution efficiency, and scalability. Finally, Section~\ref{sec:conclusions} summarize the key findings, emphasizing ORR’s security advantages without losing QoS confirming its feasibility against the KR and TN models.

\section{Related work} \label{sec:related}

The two main approaches to counter distance limitations in QKDNs, assuming commercial quantum repeaters are unavailable, are KR and TN. These approaches pose security risks, as they require all nodes to be trusted~\cite{huttner2022}. Rass \textit{et al.}~\cite{rass2024} demonstrated that in such networks, security is bounded by the least trusted intermediate node. To address this, De Santis \textit{et al.}~\cite{desantis2024} explored satellite-based communication to reduce reliance on multiple intermediate nodes, instead focusing on configurations with only one or two trusted intermediaries, an assumption feasible only in satellite scenarios.

Calsi \textit{et al.}~\cite{calsi2025} propose a novel method for KR networks, where each node establishes QKD channels with both its nearest and next-nearest neighbors. While this reduces the risk of single-node attacks, it remains vulnerable to scenarios with multiple ($n>1$) malicious nodes. In TN networks, Vyas \textit{et al.}~\cite{vyas2024} propose trust-level segmentation to relax security conditions at intermediate nodes, allowing for more flexible operations. This approach, while promising, requires additional network mechanisms to establish trust levels reliably.

Malicious nodes represent an inevitable security concern in non-repeater quantum networks, requiring for additional cryptographic defenses. PQC has been integrated into classical communication protocols~\cite{schwabe2020, shim2025} to mitigate security risks posed by emerging commercial QPUs. Rios \textit{et al.}~\cite{rios2025} demonstrate how Kyber and Dilithium improve the performance of classical cryptographic algorithms at high-security levels, albeit at the cost of increased network traffic.

For both short-term and long-term security, hybrid solutions combining QKD and PQC are emerging as the preferred approach for critical infrastructure~\cite{zeng2024a} in the NISQ era, providing the highest levels of security~\cite{wang2021, zeng2024b}. One such approach involves using QKD-generated keys to encrypt the key agreement process with PQC, as shown in~\cite{djordjevic2020, wang2021, rani2025}.

Given this scenario, we adopt a hybrid solution, combining QKD in consecutive nodes in a KR manner with OR and PQC encryption, to share a secret key between distant nodes.

\section{Technological background} \label{sec: tech-bg}

The security of QKD and PQC processes is guaranteed by distinct sources, and thus the mechanisms to individually break the security are also different. Interestingly, this makes for a valuable trade-off: for maximizing point-to-point security QKD is the best option, while for maximizing throughput PQC overcomes QKD, both aiming for the highest achievable security. One can instead combine the two processes to incorporate both of its strengths and achieve and unparalleled level of communication security, for short and long term, against any attacker. Our approach involves using OR with PQC encrypted keys in a KR network to create multiple layers of security, ensuring that 
point-to-point security is unconditional.

\subsection{Key-relay quantum key distribution networks}
In the KR approach, two nodes that wish to share a common secret ($S$) first determine the optimal path between them, i.e., the path with the least number of intermediate QKD nodes. The sender node then generates $S$ using a quantum random number generator (QRNG), encrypts it with the quantum key shared with its neighbor and transmits it over a classical channel. Each intermediate node decrypts the incoming ciphertext and re-initiates this process until the destination node recovers $S$. The main drawback of this method is that all intermediate nodes have access to $S$, which compromises the confidentiality of the shared key.

It is also worth mentioning the potential for spoofing attacks on the classical channel in KR implementations. Such attacks may be initiated by any node. While it is unlikely that an attacker would obtain the quantum key necessary to decrypt the messages, since it would have to take control of a QKD node in order to take the quantum keys, the possibility of network disruption remains a concern.

\subsection{Onion routing}\label{sec:onion}

OR is a method of anonymous communication designed to protect the identity and privacy of users on the Internet. Its operation is based on layered encryption, where a message is encrypted multiple times. Each node in the network only knows the previous and next node in the path, which prevents any single entity from having access to the complete sender and receiver information. The process begins with the selection of a path of intermediate nodes, after which the message is encrypted in successive layers, starting with the public key of the last node and progressing to the first. As the message traverses the network, each node decrypts its corresponding layer with its asymmetric key and forwards the remaining content to the next node, until finally the message reaches the exit node, where the last layer of encryption is removed and delivered to the final destination. This approach offers a high level of anonymity, since no intermediate node can know both the source and destination of the message.

OR has a large number of variants and extensions that improve security by adapting the protocol to specific scenarios. Among the most interesting extensions when sending messages are those that implement integrity~\cite{kuhn2020,kuhn2021} between the initiator of the circuit and the onion routers. Thanks to integrity, the destination node can guarantee that the incoming message has not been tampered with and can prevent attacks such as message forwarding that can cause overloads on onion routers or even disable them. Nonetheless, the implementation of OR introduces certain limitations, including increased communication latency and the potential vulnerability of outgoing nodes to surveillance or attacks.

\subsection{Kyber and Dilithium}
PQC algorithms are designed to resist quantum attacks. Currently, most vulnerabilities in these protocols stem from implementation errors, as PQC is still undergoing standardization. To address this, NIST is leading an initiative to standardize various PQC algorithm families. This work focuses on two algorithms that have successfully passed multiple selection rounds: Kyber and Dilithium.

Kyber is a lattice-based key encapsulation mechanism (KEM) used to securely establish symmetric keys over a public channel. Its security is based on the Module Learning With Errors (MLWE) problem~\cite{langlois2014}, which is currently believed to be resistant to quantum attacks. In this work, we utilize the Kyber-768 variant.

Key establishment in Kyber proceeds as follows:
\begin{enumerate}
    \item Key generation: A secret key $S$ and a public key $A$ are generated. The public key $A$ is shared, while the secret key $S$ remains private at the receiver.
    \item Encapsulation: The sender uses $A$ to encapsulate a randomly generated shared secret $C$, producing a ciphertext.
    \item Decapsulation: The ciphertext is transmitted to the receiver, who then uses their private key $S$ to decapsulate and recover $C$.
\end{enumerate}

On the other hand, Dilithium is a lattice-based signature algorithm, used to provide authentication and integrity in quantum resistant networks. Its security is based in MLWE and also on a variant of Module Short Integer Solution (MSIS) problems called Self Target MSIS~\cite{kiltz2018}.

In this case, the signature process is:
\begin{enumerate}
    \item Key generation: A couple of secret vectors $S_1$ and $S_2$ are used to generate a public key $A$. The public key $A$ is shared, while the secret vectors remain private at the receiver.
    \item Signing: The receiver applies a hash function to the message $m$, and generates a signature $Q$ using $S_1$ and the hashed messages, which are all sent to the sender as the signed message.
    \item Verification: Now the sender with the signed message and $A$ can verify the signature and if it is accepted communication can be established.
\end{enumerate}

\section{Model Principles} \label{sec:model}

The previous section presented all the individual mechanisms behind the evaluated model, all with their respective  advantages and disadvantages. For QKDN to be unconditionally secure, all of them must be combined to ensure CIA principles and anonymity. To address the confidentiality issues that arise in key sharing within QKDN security mechanisms in the KR and trusted node models, a key distribution model inspired by Onion Routing Protocol (OR) together with the KR model, referred to as OR-Relay (ORR), is proposed in this work. OR encryption with a PQC-KEM will provide confidentiality as the secret key will only be available in the extremes of the communication, while simultaneously providing anonymity. Integrity and authenticity can be guaranteed node-to-node thanks to QKD, and end-to-end is covered by the OR protocol presented in~\cite{kuhn2019}.

ORR,  as stated before, aims to ensure confidentiality in the transmission of a secret within a QKDN by utilizing cryptographic material obtained through QKD and PQC-KEM methods, combined with layered encryption. In this study, integrity and message signing mechanisms are omitted since they fall outside the scope of this paper. Assuming that neighboring nodes already share a QKD key, the basic operation of the ORR model is as follows:

\begin{figure*}[t]
    \centering
    \includegraphics[width=\linewidth]{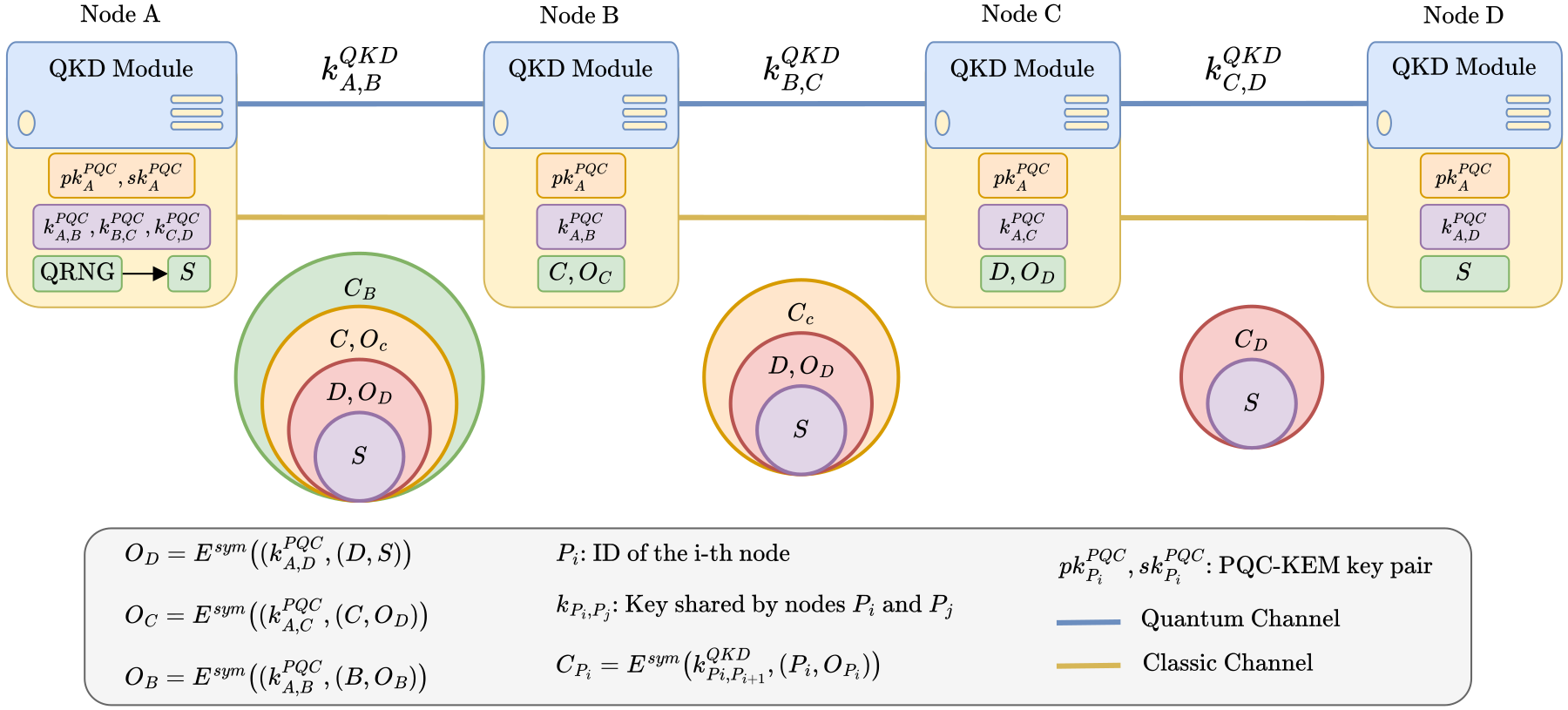}
    \caption{Simplified example of the Onion Routing Relay key distribution model}
    \label{fig:ORR}
\end{figure*}

\begin{itemize}
    \item The initial QKD node \(N_i\) creates a circuit that includes the destination QKD node \(N_d\), the receiver of the secret \(S\) that is going to be shared, as well as the \(N\) intermediate QKD nodes \(N_{\text{int}}\) required within the QKDN to establish communication between \(N_i\) and \(N_d\).
    \item \(N_i\) initiates a negotiation process with each \(N_{\text{int}}\) and \(N_d\) using a PQC-KEM to establish a symmetric key \(K^{PQC}_{sim}\) with each node in the circuit.
    \item \(N_i\) generates a random secret using a Quantum Random Number Generator (QRNG).
    \item \(N_i\) begins layered encryption\footnote{Encapsulated ciphertexts in layered encryption are commonly referred to as onions}, starting with the \(K^{PQC}_{sim}\) of the farthest node (\(N_d\)). First, \(S\) is encrypted with \(K^{PQC}_{sim}(N_d\), and the result is then encrypted with \(K^{PQC}_{sim}(N_{\text{int},N})\). This process is repeated for the remaining intermediate nodes until encryption is completed using \(K^{PQC}_{sim}(N_{\text{int},1})\).
    \item Once \(N_i\) obtains the final onion, it is encrypted with the QKD key shared between \(N_i\) and \(N_{\text{int},1}\) and then transmitted to the latter.
    \item \(N_{\text{int},1}\) decrypts the ciphertext using the QKD key shared with \(N_i\), obtaining the onion, which it is then decrypted using its own \(K^{PQC}_{sim}(N_{\text{int},1})\) before forwarding the result to the next intermediate node.
    \item This process is repeated across the remaining intermediate nodes until it reaches \(N_d\), which, upon decryption, will not encounter another onion but instead the shared secret \(S\).
\end{itemize}

Figure~\ref{fig:ORR} illustrates an example of the encryption operation in the ORR model. It should be noted that in the original OR protocol the layered encryption to create onions is done directly with the public keys of the other nodes. However, since PQC public key cryptography has been developed for the creation of KEMs, it is not possible to encrypt large amounts of information and it is limited to share symmetric keys that are used in ORR for the creation of onions. 

Regarding the symmetric encryption algorithm used, any algorithm considered secure, even taking into account the feasibility of Grover’s algorithm, would be valid. For example, AES-128 is currently deemed secure; therefore, to ensure the same level of security, AES-256 should be used to be resistant to quantum threats.

\subsection{OR-Relay vs KR \& TN models}

Once the basic operation of ORR has been defined, it is time to compare it theoretically with the alternatives it aims to improve: KR and TN. The main advantage of ORR is the security it offers over KR and TN by preventing the secret to be shared from being discovered by a malicious intermediate node. Nevertheless, the encryption mechanism used is more complex and slower than KR and TN, which only use their QKD keys to encrypt the secret or other keys using the XOR operation, which guarantees ITS and higher execution speed. 

In the scenario where there is a malicious node $N_m$ in the QKDN, for the case of the KR model it gets the key directly at the moment of decrypting the ciphertext coming from its previous node without any additional effort. In the case of the TN, if $N_m$ manages to get hold of the encrypted messages sent by the rest of the nodes through the classical channel to TN, $N_m$ can also get hold of $S$ by XOR properties. Finally, for ORR, even if $N_m$ tries to discover $S$, it will not be able to do so because it will be protected by the onion with the symmetric PQC key of the next node guaranteeing confidentiality. In fact, even if the PQC-KEM is compromised and the symmetric key is obtained, the security level is equivalent to the one obtained in the KR model.

In addition to confidentiality, ORR also provides anonymity in the transmission of the secret in order to prevent the identity of the destination node from being revealed. Anonymity was the main design feature of the OR protocol inherited by the ORR model.

On the other hand, the increase of encryption in ORR is associated with processing time that can result in loss of the final QoS, i.e., the time it takes for the secret to reach the destination node since it is generated. Both KR and TN are designed to use the XOR operation for the encryption of their secrets, making the encryption much faster than other types of algorithms such as AES. From this point of view, the KR model is the one that requires the least amount of encryption time, followed by the TN which adds the processing at the trusted node to create the message to be sent to the destination node. Finally, ORR with its layered PQC key encryption and QKD key encryption is the one that will require more encryption time. 

Furthermore, it is important to take into account the message transmission time between the QKD nodes, and the TN in its model, through the classical channels. Without considering the negotiation and management messages, KR and ORR send the same number of messages, the latter being larger due to the type of onion encryption. Meanwhile, in the TN model, one more message is added to the communication: All nodes except the destination send a ciphertext to the TN and the latter sends the final encryption to the destination node. The advantage of the TN model is that all messages received by the TN can be send concurrently, not as in KR and ORR where the sending is sequential, so the key distribution time can be better or worse depending not only on the model but also on the scenario. 

In the next section, the performance of ORR versus KR and TN in a realistic scenario is discussed, taking into account both encryption time and key distribution time to evaluate the feasibility of ORR in a real QKDN.

\section{Tests performed \& Implementation} \label{sec:tests}

To compare the different models, a series of scripts were developed in C to simulate their behavior. A GitHub repository\footnote{\url{https://github.com/pedrotega/ORRvsTNvsKR}} is available to readers, providing the source code and execution instructions.

For the experiment, only a pair of QKD nodes was available, specifically the Cerberis QKD XGR~\cite{idq2021} developed by ID Quantique, which represents the QKDN. Each simulated node in the scripts requests QKD keys from one of the real nodes, which works as a KMS, and retrieves the key by its identifier from the other real node. All the request follows the RESTful API format specified in ETSI-014~\cite{etsi2019}.

In the code, each node is executed with a different thread that allows the messages and ciphertexts of the different models to be transmitted between the nodes through the use of shared variables. In addition, the threaded approach also allows the execution of threads concurrently or sequentially, as appropriate, at any given time.

Since a QRNG was not available for this project, a PRNG (Pseudo-RNG) had to be implemented as a substitute. Table~\ref{tab:crypto_algorithms} shows the algorithms or functions that have been used in the implementation together with the software supplier.

\begin{table}[t]
    \centering
    \caption{Cryptographic Algorithms and Their Implementations}
    \begin{tabular}{|c|c|c|}
        \hline
        \textbf{Category} & \textbf{Algorithm/Function} & \textbf{Implementation} \\ 
        \hline
        Symmetric Encryption & AES-256-CBC & OpenSSL~\cite{openssl} \\
        \hline
        PQC-KEM & ML-KEM (Kyber-768) & OQS~\cite{oqs} \\ 
        \hline
        PRNG & RAND\_bytes() & OpenSSL~\cite{openssl} \\ 
        \hline
        XOR & Directly in C & N/A \\ 
        \hline
    \end{tabular}
    \label{tab:crypto_algorithms}
\end{table}

The parameters that have been measured in the tests are the encryption time and the distribution time computed from the moment the secret to be shared is created in the initiator node until it reaches the destination node. As for the encryption time, the measurement varies depending on the model:

\begin{itemize}
    \item \textbf{Onion Routing Relay}: The longest encryption time in this model is spent encrypting the initial onion at the initiator node.
    
    \item \textbf{Trusted Node}: In this case, the processing time of the encrypted messages within the trusted node is measured when calculating the ciphertext to be sent to the destination node.
    
    \item \textbf{Key Relay}: Being the simplest alternative, it is only necessary to measure the XOR operation when encrypting the secret in the initiator node.
\end{itemize}

\section{Results} \label{sec:results}

\begin{figure}[t]
    \centering
    \includegraphics[width=\linewidth]{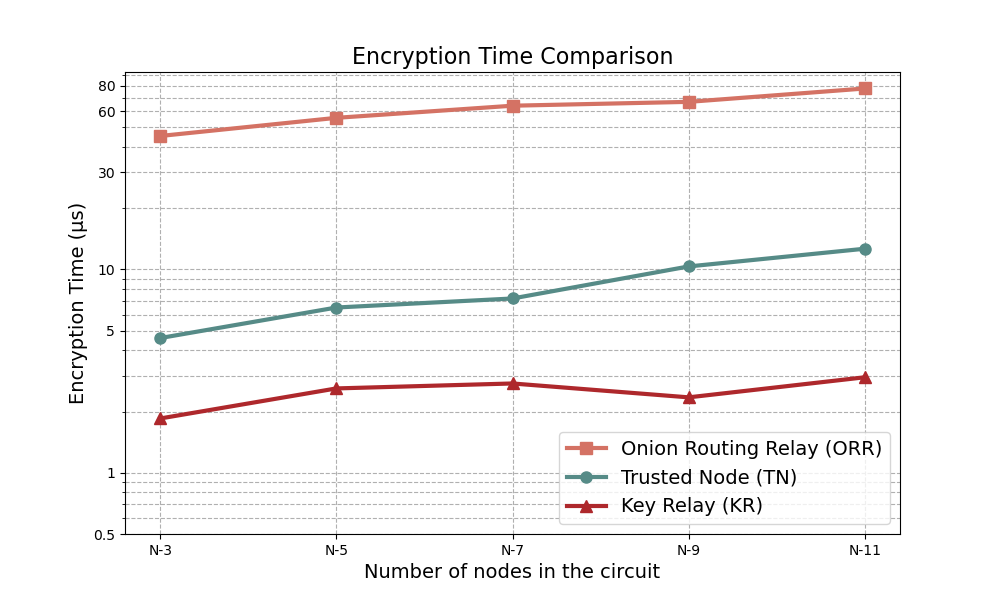}
    \caption{Encryption time of the different models}
    \label{fig:enc-time}
\end{figure}

This section presents the results of the measurements conducted in the test scenario. Figure~\ref{fig:enc-time} illustrates the average time required for the encryption procedure in each model. The KR model exhibits a consistent encryption time ranging from 1,5 to 3 $\mu s$, which remains largely unaffected by the number of nodes, since its encryption is the XOR operation of the secret with a QKD key regardless of the number of nodes. In contrast, the TN model shows an average encryption time of 4,6 $\mu s$ for a 3-node circuit, which increases to 12,5 $\mu s$ as the number of ciphertexts received from the other nodes increases. For the ORR model, the implementation of layered encryption results in a significant increment in time, with an average of 45,35 $\mu s$ for 3-node circuit and 77,65 $\mu s$ when the network consists of 11-node circuit\footnote{In all the tests performed, the destination node is inside the circuit.}.

\begin{figure}[t]
    \centering
    \includegraphics[width=\linewidth]{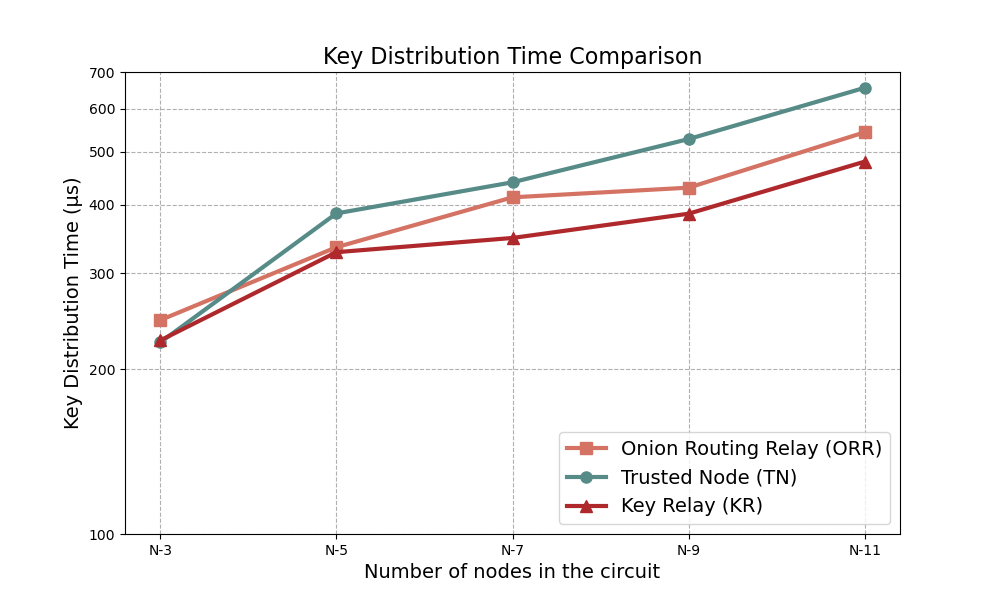}
    \caption{Key distribution time of the different models}
    \label{fig:dist-time}
\end{figure}

Regarding the time measurement of the key distribution, which is represented in Figure~\ref{fig:dist-time}, it can be seen that the time obviously increases with the number of nodes. Again KR remains as the fastest model, in general, starting with 226 $\mu s$ for a 3-node circuit and reaching its maximum with 11 nodes and almost surpassing 480 $\mu s$. In the case of ORR it starts as the slowest alternative with about 246 $\mu s$ but consolidates as the second fastest option taking for the 11-node circuit 541 $\mu s$ in average. On the other hand, the TN model is the one that experiences the greatest growth, needing even less time than KR for a 3-node circuit (224 $\mu s$) but reaching 655 $\mu s$ for the 11-node circuit. It should be noted that, in addition to the threads for the QKD nodes, this option also requires another thread to act as TN.

\section{Conclusions \& Future work} \label{sec:conclusions}

From the results obtained in the previous section it can be concluded that, as predicted, ORR layered encryption requires between 10 to 40 times more $\mu s$ compared to the encryption processes of the TN and KR models respectively. Moreover, the increase in time of the TN with respect to KR is due to the processing to be done at the trusted node which increments along with the number of nodes.

Nonetheless, the most notable difference is given by the key distribution time, which is the one that ultimately affects the system's QoS. In the results, it is found that the context switch time between threads, which simulates the transmission time between nodes, increases the distribution time the most. In a real scenario, this interval will become more remarkable as the distance between nodes grows and thereby also the transmission time. This characteristic may explain why the TN model exhibits the longest encryption time for this parameter. Although messages containing cryptographic material can be transmitted concurrently, the model also requires the transmission of an additional message to the destination node, contributing to the increased key distribution time. Conversely, while the layered encryption significantly increases the encryption time in the ORR model, the overall key distribution time is altered by an order of magnitude, resulting in a final performance comparable to the one of the KR model. For instance, in the tests with 11 nodes, the average time difference between the KR and ORR models is approximately 60 $\mu s$, which is similar to the difference observed in their encryption times.

Therefore, it can be concluded that in a real scenario in which the message sending time is likely to increase from $\mu s$ to $ms$ and the encryption time of the models will be constant, ORR can have a competitive QoS with respect to its alternatives and improve security by preventing the disclosure of the secret to malicious nodes.

In future lines of work, it is intended to implement the integrity and authenticity mechanisms of the ORR model to check if it maintains an acceptable QoS in relation to the KR and TN versions that add authenticity through the use of PQC sign algorithms.

\end{document}